\documentclass[12pt]{article}

\begin{document}

\begin{center}
{\bf A note on generalized electrodynamics } \\
\vspace{5mm} S. I. Kruglov\\
%\footnote{E-mail:krouglov@utsc.utoronto.ca}
 \vspace{3mm} \textit{University of Toronto at Scarborough,\\ Physical
and Environmental Sciences Department, \\
1265 Military Trail, Toronto, Ontario, Canada M1C 1A4} \\

\vspace{5mm}
\end{center}

\begin{abstract}
The generalized Maxwell equations with arbitrary gauge parameter are
considered in the $11\times 11$-matrix form. The gauge
invariance of such a model is broken due to the presence of a scalar field.
The canonical and symmetrical Belinfante energy-momentum tensors are found.
The dilatation current is obtained and we demonstrate that the theory possesses the dilatation
symmetry. The matrix Schr\"{o}dinger form of equations is derived. The non-minimal interaction
in curved space-time is introduced and equations are considered in Friedmann$-$Robertson$-$ Walker
background. We obtain some solutions of equations for the vector field.
\end{abstract}

\section{Introduction}

We have investigated the first-order formulation of the generalized Maxwell equations
which describe massless vector fields with an additional scalar field in \cite{Kruglov}, \cite{Kruglov1}
(see also \cite{Kruglov2}). Such a model is not a gauge-invariant and can be treated as a Maxwell theory in the definite gauge. Gradient terms were introduced in Maxwell equations by many authors (see references in \cite{Kruglov2} and \cite{Dvoeglazov}). Here, we take into account a gauge parameter which allows us to consider different gauges. It should be mentioned that gauge parameter is physical value in our scheme, contrarily to classical electrodynamics, which can contribute to gravity interaction. Therefore, schemes with difference gauges are not equivalent each other. As was mentioned in \cite{Kruglov1} the reason for leaving a scalar field in the spectrum is the application of such a
non-gauge-invariant model in astrophysics. We have stressed \cite{Kruglov1} that the additional degree, a scalar field,
can play an important role in the inflation theory of universe. Later, authors of the work \cite{Jose} considered a
scalar field of generalized electrodynamics as a source of dark energy. Dark energy is introduced to explain
the acceleration of expanded universe at the present time and results in the negative pressure. Dark energy interacts only gravitationally representing weakly coupled substance. Scenario of inflationary universe \cite{Linde} allows us to understand observable data: our universe is homogeneous and isotropic for scales $>100$ Mpc ($1$ pc = $3.26$ light years) and expands in accordance with the Hubble law. In the chaotic inflation model a massive scalar field (quintessence) minimally coupled to gravity is responsible for slow-roll inflation and plays the role of dark energy. But, in this model, the potential terms should be fine tuned to have the acceleration of universe at the definite time. Another phenomenological way to describe dark energy is to introduce a cosmological constant $\Lambda$ into the Einstein equation (the term ($-\Lambda g_{\mu\nu}$) in the left side of the Einstein equation). One implies the existence of vacuum energy by introducing the cosmological constant. If $\Lambda>0$ the additional cosmological term leads to anti-gravity. But in this case the difficulty arises: a vacuum solution is not a Minkowski space-time. In addition, there is no physical explanation of a coincidence problem: matter and dark energy densities possess the same orders of values at the present time and had big difference in magnitudes in previous eras. Today, approximately 70$\%$ of the energy density of the universe is in the form of dark energy, and the rest 30$\%$ is in the form of non-relativistic matter. Cosmological constant, which gives the energy density, remains constant during the expansion of the universe, but the energy density of matter and radiation decreases in time. Thus, the nature of dark energy is one of the most important problems in astrophysics.

In the scenario suggested in \cite{Jose}, the time component of a field in generalized electrodynamics grows in time and becomes dominant explaining the acceleration of universe. Therefore, it is of great interest further investigation of the generalized electrodynamics where the addition degree, the scalar state of the field, can play the role of dark energy.

The paper is organized as follows. In Sec.2, the generalized Maxwell equations with arbitrary gauge parameter
are formulated in the matrix form. Matrices of the relativistic wave equation (RWE) obey the generalized
Duffin $-$Kemmer$-$Petiau (DKP) algebra. We obtain, in Sec.3, canonical and the Belinfante dilatation currents which are not conserved. The conserved modified dilatation current is also found demonstrating the scale invariance of the theory of massless fields. We obtain the Schr\"{o}dinger form of equations and the quantum-mechanical
Hamiltonian in Sec.4. The minimal equation for the matrix Hamiltonian is found. In Sec.5, a novel non-minimal interaction in curved space-time is introduced and equations are considered in Friedmann$-$Robertson$-$Walker (FRW) background. The solution of equations for the time component of the four-potential is found which grows in time. We discuss the results obtained in Sec.6. The quantum-mechanical Hamiltonian is found from relativistic wave equation in Appendix A. Starting with the second-order formulation of the theory, we obtain canonical and symmetrical energy-momentum tensors and dilatation currents in Appendix B. In Appendix C the quantization of fields is performed in the second-order formalism.

The Euclidean metric is used in Sec.1-4 and Appendixes A, B and C, and four-vectors are $x_\mu=(x_m,x_4)=(x_m,ix_0)$, and $x_0$ is a time; Greek letters run $1,2,3,4$ and Latin letters run $1,2,3$. We use natural units $\hbar = c = 1$.

\section{First-order form of equations}

In \cite{Kruglov3}, we considered the general Lagrangian form of massive vector fields. For the case of neutral massless vector fields it reduces to
\begin{equation}
{\cal L} =\delta_{\mu\nu,\sigma\rho}\left( \partial_\mu A_\sigma\right)\left( \partial_\nu A_\rho\right) ,
\label{1}
\end{equation}
where $\delta_{\mu\nu,\sigma\rho}=a\delta_{\mu\nu}\delta_{\sigma\rho}+b\delta_{\mu\sigma}\delta_{\nu\rho}
+c\delta_{\mu\rho}\delta_{\sigma\nu}$. To have the standard kinetic term, we put $a=-1/2$. The Euler-Lagrange equations
follow from (1):
\begin{equation}
\partial_\mu^2A_\nu-2\left(b+c\right)\partial_\nu\partial_\mu A_\mu=0.
\label{2}
\end{equation}
One can see from Eq.(2) that only one parameter ($b+c$) remains in the equations of motion.
It is convenient to choose: $c=1/2$, $2b=-\xi$, where $\xi$ defines the gauge. Then Eq.(1) and (2) become
\begin{equation}
{\cal L} =-\frac{1}{4}F_{\mu\nu}^2-\frac{1}{2}\xi\left( \partial_\nu A_\nu\right)^2 ,
\label{3}
\end{equation}
\begin{equation}
\partial_\nu F_{\mu\nu}-\xi\partial_\mu\partial_\nu A_\nu=0,
\label{4}
\end{equation}
where $F_{\mu\nu}=\partial_\mu A_\nu-\partial_\nu A_\mu$ is the field strength.
Eq.(4) can be treated as the Maxwell equations with the additional gauge parameter. In QED the physical values do not depend on the gauge \cite{Faddeev}, but in our scheme, we expect the dependence on $\xi$ because the scalar state presents in the spectrum. In \cite{Kruglov}, \cite{Kruglov1}, we have chosen the gauge $\xi=1$. At $\xi=0$, one arrives at standard Maxwell equations. Here we imply that $\xi\neq 0$. Introducing notations $\psi _{[\mu \nu ]}=(1/\kappa)F_{\mu\nu}$,
$\psi_\mu=A_\mu$, $\psi _0=-(\xi/\kappa)\partial_\nu A_\nu$, where $\kappa$ is the mass parameter, second order equation (4) can be represented as a system of the first-order equations:
\[
\partial _\nu \psi _{[\mu \nu ]}+\partial _\mu \psi _0=0,
\]
\begin{equation}
\partial _\nu \psi _\mu -\partial _\mu \psi _\nu +\kappa\psi _{[\mu \nu ]}=0,
\label{5}
\end{equation}
\[
\partial _\mu \psi_\mu +\frac{\kappa}{\xi}\psi _0=0.
\]
We note that fields $\psi_A$ ($A=0$, $\mu$, $[\mu\nu]$) have the same dimension.
Introducing wave function $\Psi (x)=\left\{ \psi _A(x)\right\}$, and using
elements of the entire matrix algebra $\varepsilon ^{A,B}$
obeying equations: $\left( \varepsilon
^{A,B}\right)_{CD}=\delta_{AC}\delta_{BD}$, $\varepsilon
^{A,B}\varepsilon ^{C,D}=\delta _{BC}\varepsilon ^{A,D},$ Eq.(5) can be written in the first-order matrix form
\begin{equation}
\left[ \alpha _\nu \partial _\nu +\kappa \left(\frac{1}{\xi}P_s+P_t\right)\right] \Psi (x)=0,
\label{6}
\end{equation}
where
\[
\alpha _\mu =\beta _\mu ^{(1)}+\beta _\mu ^{(0)},~~\beta _\mu
^{(1)}=\varepsilon ^{\nu ,[\nu \mu ]}+\varepsilon ^{[\nu \mu ],\nu
},~~\beta _\mu ^{(0)}=\varepsilon ^{\mu ,0}+\varepsilon ^{0,\mu },
\]
\vspace{-8mm}
\begin{equation}
\label{7}
\end{equation}
\vspace{-8mm}
\[
P_s=\varepsilon ^{0,0},~~~~P_t=\frac 12\varepsilon ^{[\mu \nu ],[\mu \nu]}.
\]
At the Feynman gauge $\xi=1$, Eq.(6) is simplified because $P_s+P_t=\varepsilon ^{0,0}+(1/2)\varepsilon ^{[\mu \nu ],[\mu \nu]}$ is the projection operator but $(1/\xi)P_s+P_t$ is not. The $11\times11$ Hermitian matrices $\alpha_\mu$ obey the generalized Duffin$-$Kemmer$-$Petiau algebra \cite{Kruglov}:
\[
\alpha _\mu \alpha _\nu \alpha _\alpha +\alpha _\alpha \alpha _\nu
\alpha _\mu +\alpha _\mu \alpha _\alpha \alpha _\nu +\alpha _\nu
\alpha _\alpha \alpha _\mu +\alpha _\nu \alpha _\mu \alpha _\alpha
+\alpha _\alpha \alpha _\mu \alpha _\nu =
\]
\vspace{-8mm}
\begin{equation}
\label{8}
\end{equation}
\vspace{-8mm}
\[
 =2\left( \delta _{\mu \nu }\alpha _\alpha +\delta
_{\alpha \nu }\alpha _\mu +\delta _{\mu \alpha }\alpha _\nu
\right),
\]
and $P_s$, $P_t$ are the projection matrices, $P_s^2=P_s$, $P_t^2=P_t$, and extract the scalar ant tensor parts of the wave function, respectively. One can verify the relations:
\[
P_s\beta _\mu ^{(0)}+\beta _\mu ^{(0)}P_s=\beta _\mu ^{(0)},~~~~~~P_t\beta _\mu ^{(1)}+\beta _\mu ^{(1)}P_t=\beta _\mu ^{(1)},
\]
\[
P_s\beta _\mu ^{(1)}=\beta _\mu ^{(1)}P_s=0,~~~~~~P_t\beta _\mu ^{(0)}=\beta _\mu ^{(0)}P_t=0.
\]
Introducing the Hermitianizing matrix $\eta$ \cite{Kruglov}:
\begin{equation}
\eta =-\varepsilon ^{0,0}+\varepsilon ^{m,m}-\varepsilon ^{4,4}
+\varepsilon ^{[m4],[m4]}-\frac{1}{2}\varepsilon ^{[mn],[mn]},
\label{9}
\end{equation}
the ``conjugated" equation reads
\begin{equation}
\overline{\Psi }(x)\left[ \alpha _\nu \overleftarrow{\partial}_\nu -\kappa \left(\frac{1}{\xi}P_s+P_t\right)\right] =0,
\label{10}
\end{equation}
where $ \overline{\Psi }=\Psi ^{+}\eta =\left(-\psi_0,\psi_\mu,-\psi_{[\mu\nu]}\right)$.
Matrices $\alpha_\mu$ and $\eta$ satisfy equations: $\eta \alpha _m=-\alpha _m^{+}\eta^+$,
$\eta \alpha _4=\alpha _4^{+}\eta^+ $.
In the first-order formalism the Lagrangian can be written as follows:
\begin{equation}
{\cal L}=-\overline{\Psi }(x)\left[ \alpha _\nu \partial _\nu +\kappa \left(\frac{1}{\xi}P_s+P_t\right)\right] \Psi (x).
\label{11}
\end{equation}
Eq.(6) follows from Lagrangian (11) by varying the corresponding action. Similar to the Dirac theory,
Lagrangian (11) vanishes for fields $\Psi$, obeying RWE (6). It should be noted that in the second-order formalism, based on the equations (4), Lagrangian (3) vanishes only within four-divergence.
Lagrangian (11), with the help of Eq.(7) becomes
\begin{equation}
{\cal L}=\psi_0 \partial _\mu\psi_\mu -\psi_\mu\partial _\mu
\psi_0 -\psi_\rho\partial _\mu\psi_{[\rho\mu]}+
\psi_{[\rho\mu]}\partial _\mu\psi_\rho +\kappa\left(\frac{1}{\xi}\psi_0^2
+\frac{1}{2}\psi_{[\rho\mu]}^2\right). \label{12}
\end{equation}
One may verify that Lagrangian (12) vanishes for fields obeying equations of motion (5) and within four-divergence,
which does not influence on the equations of motion, can be represented as
\begin{equation}
{\cal L}=-\frac{1}{2\kappa}\left(\partial_\mu \psi_\nu-\partial_\nu \psi_\mu\right)^2-
\frac{\xi}{\kappa}\left(\partial_\mu \psi_\mu\right)^2. \label{13}
\end{equation}
As $\kappa$ has the dimension of the mass, we need to renormalize the fields, and under the replacement
$\psi_\mu\rightarrow (\sqrt{\kappa}/\sqrt{2})A_\mu$, Lagrangian (13) coincides with (3). Of course, one could define fields $\psi_A$ according to this replacement from the very beginning. It was pointed on the importance of normalization in \cite{Dvoeglazov1}.

\section{Energy-momentum tensors and dilatation currents}

Now, we investigate the scale invariance in the model with arbitrary gauge parameter $\xi$.
For this purpose, one needs to obtain energy-momentum tensors and dilatation currents.
The conserved canonical energy-momentum tensor (see \cite{Kruglov1}) is
\[
T^c_{\mu\nu}=\left(\partial_\nu \overline{\Psi} (x)\right)\alpha_\mu
\Psi (x)
\]
\vspace{-8mm}
\begin{equation}
\label{14}
\end{equation}
\vspace{-8mm}
\[
=\psi_0\partial_\nu \psi_\mu-\psi_\mu\partial_\nu
\psi_0-\psi_\rho\partial_\nu \psi_{[\rho\mu]} +\psi_{[\rho\mu]}
\partial_\nu \psi_\rho,
\]
so that $\partial_\mu T^c_{\mu\nu}=0$. We took here into account that Lagrangian (11) vanishes on the solutions of equations of motion. The canonical dilatation current \cite{Coleman} becomes
\begin{equation}
D_\mu^c=x_\alpha T_{\mu\alpha}^{c},
\label{15}
\end{equation}
with its non-zero four-divergence
\begin{equation}
\partial_\mu D_\mu^c=T_{\mu\mu}^{c}=\kappa\overline{\Psi }\left(\frac{1}{\xi}P_s+P_t\right)\Psi=-\kappa\left(\frac{1}{\xi}\psi_0^2 +
\frac{1}{2}\psi_{[\mu\nu]}^2\right).
\label{16}
\end{equation}
The appearance of the gauge parameter $\xi$ here is due to using equations of motion (5).
The conserved symmetric Belinfante energy-momentum tensor is given by \cite{Kruglov1}
\[
T_{\mu\alpha}^{B}=2\kappa\psi_{[\lambda\mu]}\psi_{[\alpha\lambda]}-2\psi_\mu\partial_\alpha\psi_0
-2\psi_\alpha\partial_\mu\psi_0
\]
\vspace{-8mm}
\begin{equation}
\label{17}
\end{equation}
\vspace{-8mm}
\[
+\delta_{\alpha\mu}\partial_\beta\left(\psi_0\psi_\beta\right)
-\delta_{\alpha\mu}\partial_\beta\left(\psi_\lambda\psi_{[\lambda\beta]}\right).
\]
We find non-zero trace of the symmetric Belinfante energy-momentum tensor (17):
\begin{equation}
T_{\mu\mu}^{B}=4\partial_\mu\left(\psi_0\psi_\mu\right).
\label{18}
\end{equation}
A modified dilatation current \cite{Kruglov1} is given as follows:
 \begin{equation}
D_\mu^B=x_\alpha T_{\mu\alpha}^{B}+\psi_\lambda\psi_{[\lambda\mu]}-3\psi_0\psi_\mu.
\label{19}
\end{equation}
One may verify that the divergence of the Belinfante dilatation current (19) coincides with
the divergence of the canonical dilatation current (16), $\partial_\mu D_\mu^B=\partial_\mu D_\mu^c\neq 0$.
As the trace of the Belinfante dilatation current (18) is a four-divergence, we may
define new conserved dilatation current
\begin{equation}
D_\mu=x_\alpha T_{\mu\alpha}^{B}-4\psi_0\psi_\mu,
\label{20}
\end{equation}
and $\partial_\mu D_\mu=0$.
Thus, new dilatation current (20) is conserved, and strictly speaking, the dilatation symmetry is not broken and the
model possesses the scale invariance. We have change the conclusion made in \cite{Kruglov1} about the scale invariance because of obtaining new conserved dilatation current (20).
Relations found in this section are the generalization of
formulas obtained in \cite{Kruglov1} on the case of arbitrary gauge $\xi$.

\section{Schr\"{o}dinger form of equations}

It should be noted that in some cases the Schr\"{o}dinger equation has advantages for
the investigation of interacting field problems.
To obtain the Schr\"{o}dinger form of equations and quantum mechanical Hamiltonian,
we have to exclude the non-dynamical components from Eq.(5).
For this purpose, Eq.(5) can be represented as follows:
\[
\kappa\psi _{[4 m]}=\partial _4 \psi _m -\partial _m \psi _4 , ~~~~
\partial _4 \psi _{[m 4 ]}+\partial _n \psi _{[m n]}+\partial_m \psi_0=0,
\]
\begin{equation}
\psi _{[mn]}=\frac{1}{\kappa}\left(\partial _m \psi _n -\partial _n \psi _m \right), ~~~~
 \label{21}
\end{equation}
\[
\partial _n \psi _{[4n]}+\partial_4 \psi _0 = 0,~~~~\partial _4 \psi _4 +\partial_m\psi_m+\frac{\kappa}{\xi}\psi_0=0.
\]
Third equation in (21) possesses only
spatial derivatives and, therefore, $\psi _{[mn]}$ are non-dynamical (auxiliary) components.
Excluding $\psi_{[mn]}$ from
Eq.(21), we arrive at the system of equations containing only dynamical
components
\[
i\partial _t \psi _0=\partial_n\psi_{[4n]},~~~~i\partial _t \psi _m=-\kappa\psi _{[4 m]}-\partial_n \psi_4,
\]
\vspace{-8mm}
\begin{equation}
\label{22}
\end{equation}
\vspace{-8mm}
\[
i\partial _t \psi _4=\partial_m\psi_m+\frac{\kappa}{\xi}\psi_0,~~~~i\partial_t \psi_{[m 4 ]}=\frac{1}{\kappa}\left(\partial_m\partial
_n \psi _n-\partial_n^2\psi_m\right)+ \partial_m\psi _0 .
\]
Let us introduce the 8-component wave function
\begin{equation}
\Phi (x)=\left(
\begin{array}{c}
\psi_0(x)\\
\psi_\mu (x)\\
\psi_{[m 4]}(x)\\
\end{array}
\right).
\label{23}
\end{equation}
Exploring the elements of the matrix algebra, Eq.(22) can be represented as follows:
\[
i\partial_t\Phi(x)=\biggl[\kappa\left(\frac{1}{\xi}\varepsilon^{4,0}+\varepsilon^{m,[m4]}\right)
+\biggl(\varepsilon^{0,[4m]}-\varepsilon^{[4m],0}+\varepsilon^{4,m}-\varepsilon^{m,4}\biggr)\partial_m
\]
\vspace{-7mm}
\begin{equation}
\label{24}
\end{equation}
\vspace{-7mm}
\[
+\frac{1}{\kappa}\left(\varepsilon^{[m4],n}\partial_n\partial_m-\varepsilon^{[m4],m}\partial_n^2\right)\biggr]\Phi (x).
\]
Then Eq.(24) takes the Schr\"{o}dinger form:
\begin{equation}
i\partial_t\Phi(x)={\cal H}\Phi(x),
\label{25}
\end{equation}
where the Hamiltonian is given by
\[
{\cal H}=\kappa\left(\frac{1}{\xi}\varepsilon^{4,0}+\varepsilon^{m,[m4]}\right)
+\biggl(\varepsilon^{0,[4m]}-\varepsilon^{[4m],0}+\varepsilon^{4,m}-\varepsilon^{m,4}\biggr)\partial_m
\]
\vspace{-8mm}
\begin{equation}
\label{26}
\end{equation}
\vspace{-8mm}
\[
+\frac{1}{\kappa}\left(\varepsilon^{[m4],n}\partial_n\partial_m-\varepsilon^{[m4],m}\partial_n^2\right).
\]
In the momentum space the Hamiltonian becomes:
\[
{\cal H}=\kappa\left(\frac{1}{\xi}\varepsilon^{4,0}+\varepsilon^{m,[m4]}\right)
+ik_m\biggl(\varepsilon^{0,[4m]}-\varepsilon^{[4m],0}+\varepsilon^{4,m}-\varepsilon^{m,4}\biggr)
\]
\vspace{-8mm}
\begin{equation}
\label{27}
\end{equation}
\vspace{-8mm}
\[
+\frac{1}{\kappa}k_m k_n\left(\varepsilon^{[k4],k}\delta_{mn}-\varepsilon^{[m4],n}\right).
\]
The 8-component wave function (23) describes fields with four spin
states with positive and negative energies, and there are only dynamical components
in the wave function. The matrix Hamiltonian (27) obeys the minimal equation as follows:
\begin{equation}
{\cal H}^2\left({\cal H}^2- \textbf{k}^2\right)^2\left({\cal H}^2-
2\textbf{k}^2\right)=0.\label{28}
\end{equation}
There are three eigenvalues of the Hamiltonian: $0$, $\textbf{k}^2$, $2\textbf{k}^2$.
The physical eigenvalue is $\textbf{k}^2$ so that $k^2_0=\textbf{k}^2$. In Appendix A, one can find
the Hamiltonian (26) expressed through the matrices (7) of equation (6).

\section{Interaction with gravitation}
\subsection{Equations}

Now, we consider the non-minimal interaction of the massless vector field with gravity. Some aspects of interactions of vector fields with gravity where investigated in \cite{Ford}, \cite{Kiselev}, \cite{Picon},\cite{Novello}, \cite{Wei} \cite{Boehmer}, \cite{Golovnev}, \cite{Chiba}, \cite{Esposito}. Let us consider the novel action
\begin{equation}
S=\int d^4x\sqrt{-g}\left[-\frac{R}{16\pi G}-\frac{1}{4}F_{\mu\nu}F^{\mu\nu}-\frac{1}{2}\xi\left( \nabla_\nu A^\nu\right)^2+\lambda R\nabla_\nu A^\nu \right],
 \label{29}
\end{equation}
where $G$ is the gravitational (Newton) constant, $\lambda$ is a coupling constant, and $\nabla_\mu$ are covariant derivatives.
In Eq.(29), we have introduced the coupling of a scalar curvature with the vector field $A_\mu$.
Within the four-divergence, which does not change equations of motion, the non-minimal interaction term in action (29) also can be represented as
\[
S_{int}=-\lambda\int d^4x\sqrt{-g}A^\nu \frac{\partial R}{\partial x^\nu}.
\]
Such term can be added for any vector-tensor theory of gravity but it will lead to the higher derivative model.
In \cite{Jose1} other couplings to gravity were investigated. Let us consider
FRW space-time with the flat spatial part with the metric
\begin{equation}
g_{00}=g^{00}=1,~~g_{11}=g_{22}=g_{33}=-a(t)^2,~~g^{11}=g^{22}=g^{33}=-\frac{1}{a(t)^2},
 \label{30}
\end{equation}
where $a(t)$ is a scale factor.
Nonzero components of the Christoffel symbols and curvatures are given by
\[
\Gamma_{10}^1=\Gamma_{20}^1=\Gamma_{30}^3=\frac{\dot{a}}{a}=H,~~~~
\Gamma_{11}^0=\Gamma_{22}^0=\Gamma_{33}^0=\dot{a}a=a^2H,
\]
\vspace{-8mm}
\begin{equation}
\label{31}
\end{equation}
\vspace{-8mm}
\[
R_{00}=-3\frac{\ddot{a}}{a},~~~~R_{ik}=\delta_{ik}\left(a\ddot{a}+2\dot{a}^2 \right),~~
R=-6\left[\frac{\ddot{a}}{a}+\left(\frac{\dot{a}}{a}\right)^2\right],
\]
where $H=\dot{a}(t)/a(t)$ is the Hubble parameter; $R_{\mu\nu}$, $R$ are the Ricci and scalar curvatures,
respectively. Varying action (29) with respect to vector field $A_\mu$ yields the equations of motion as follows:
\begin{equation}
\nabla_\nu F^{\mu\nu}-\xi\nabla^\mu\nabla_\nu A^\nu +\lambda\nabla^\mu R=0.
\label{32}
\end{equation}
For homogeneous electromagnetic fields ($\partial_i A_\mu=0$) the $\mu=0$ component of Eq.(32) in FRW background gives
\begin{equation}
\nabla_\nu F^{0\nu}-\xi\nabla^0\nabla_\nu A^\nu +\lambda\nabla^0 R=-\xi\left(\ddot{A}^0+3\dot{H}A^0+3H\dot{A}^0\right)+\lambda\dot{R}=0.
\label{33}
\end{equation}
Taking into account Eq.(31), one obtains from Eq.(33) the equation for $A_0$:
\begin{equation}
\left[\ddot{A}_0+\partial_t\left(3HA_0-\frac{\lambda}{\xi}R\right)\right]=0.
\label{34}
\end{equation}
The equation for spatial components are given by
\begin{equation}
\ddot{A}_m+H\dot{A}_m=0.
\label{35}
\end{equation}
Variation of action (29) with respect of the metric leads to generalized Einstein's equation.

\subsection{Solutions}

Let us obtain solutions to novel equation (34). Integrating Eq.(34), one finds
\begin{equation}
\dot{A}_0+3HA_0-\frac{\lambda}{\xi}R=C_1,
\label{36}
\end{equation}
where $C_1$ is the integration constant. Eq.(36) with the help of Eq.(31) becomes the first-order non-homogeneous
differential equation:
\begin{equation}
\dot{A}_0+3\frac{\dot{a}}{a}A_0=C_1- \frac{6\lambda}{\xi}\left[\frac{\ddot{a}}{a}+\left(\frac{\dot{a}}{a}\right)^2\right].
\label{37}
\end{equation}
We can take $H = p/t$, with $p = 1/2$ for radiation and $p = 2/3$ for matter eras respectively.
In this case, we obtain the solution to Eq.(37):
\begin{equation}
A_0=C_1t +C_2t^{-3p}-\frac{6\lambda p(2p-1)}{\xi (3p-1)}t^{-1}.
\label{38}
\end{equation}
The last term in Eq.(38) is due to non-minimal interaction introduced in Eq.(29). The first term of $A_0$-component grows with the cosmic time for any $p$ \cite{Jose}, and the last term decays.
The $A_i$-component grows but the temporal component ($A_0$) dominates \cite{Jose}. As was noted in \cite{Jose}, the
term $\nabla_\mu A^\mu$ plays the role of a cosmological constant during the evolution of the universe.

\section{Conclusion}

The generalized Maxwell equations with arbitrary gauge parameter are formulated in the first-order formalism. The gauge U(1)-symmetry of a model is broken. As a result, the scalar state of the field presents in the spectrum of the theory.
If one introduces the four-current $J_\mu$ in the right side of Eq.(4), then due to the conservation of the current, $\partial_\mu J_\mu=0$, the equation $\partial_\mu^2 \partial_\nu A_\nu =0$ holds. It means that the scalar state of the field $A_\mu$ does not interact with charges and currents. But this scalar state can interact with gravity via the coupling (29), we have introduced.

As the matrices of the RWE obey the generalized DKP algebra, one can apply covariant methods for finding solutions for definite spin (one and zero), spin projections and energy-momentum \cite{Kruglov}, \cite{Kruglov1}. Although the canonical and Belinfante dilatation currents, found within the first-order formalism, are not conserved, we have obtained the conserved modified dilatation current. This demonstrates the scale invariance of the massless fields theory.
The Schr\"{o}dinger form of equations obtained possesses some advantages because it contains only dynamical components of fields. The found quantum-mechanical Hamiltonian can be used for investigation of problems with interacting fields.
RWE, as well as Hamiltonian, are simplified for the choice $\xi=1$ which was used in \cite{Kruglov}, \cite{Kruglov1}.
In addition, the commutation relations (C5) take the canonical normalized form at $\xi =1$. In our opinion the value of the parameter $\xi =1$ is natural. The consistency of the model will be studied in subsequent papers.

We suggest a novel non-minimal interaction of fields in the FRW background.
The solution of equations for the time component of the four-potential grows in time in the same manner as in \cite{Jose}. Therefore, the model considered has the similar behavior. Although there are difficulties with the unbounded Hamiltonian and indefinite metrics (see Appendix C), the model has attractive features. We leave the detailed analysis of the model based on the action (29) for further investigations.

\vspace{7mm}
\textbf{\textbf{Acknowledgments}}
\vspace{7mm}

I wish to thank V. V. Dvoeglazov for correspondence and useful discussions.
\vspace{5mm}

\textsc{\textbf{Appendix A}}

\vspace{5mm}

Let us obtain the Schr\"{o}dinger equation and quantum-mechanical Hamiltonian from relativistic wave equation (6).
One can find from (6) the equation as follows:
\[
i\alpha _4\partial _t\Psi (x)=\left[\alpha _a\partial_a+\kappa
\left(\frac{1}{\xi}P_s+P_t\right)\right]\Psi (x).~~~~~~~~~~~~~~~~~~~~~~~~~~~~~~~~~~~~~~~~~~~~~~~~~~~(A1)
\]
Taking into account the relation $\alpha _4\left(\alpha _4^2-1 \right)=0$, which follows from
algebra (8), one can introduce the projection operators:
\[
\Lambda\equiv\alpha_4^2=\varepsilon^{0,0}+\varepsilon^{\mu,\mu}+
\varepsilon^{[m4],[m4]},~~~~\Pi \equiv 1-\alpha_4^2
=\frac{1}{2}\varepsilon^{[mn],[mn]}, ~~~~~~~~~~~~~~~~~~~~~~~~~~~~(A2)
\]
with the properties $\Lambda=\Lambda^2$, $\Pi^2=\Pi$, $\Lambda\Pi=\Pi\Lambda=0$, $\Lambda+\Pi=I_{11}$
($I_{11}$ is unit $11\times11$-matrix).
Operator $\Lambda$ extracts 8-dimensional subspace of dynamical components and operator $\Pi$ acts in 3-dimensional
subspace of non-dynamical components of the wave function $\Psi$. Thus,
we introduce dynamical, $\phi(x)$, and non-dynamical, $\chi(x)$, functions:
\[
\phi(x)=\Lambda\Psi(x),~~~~\chi(x)=\Pi \Psi(x). ~~~~~~~~~~~~~~~~~~~~~~~~~~~~~~~~~~~~~~~~~~~~~~~~~~~~~~~(A3)
\]
After multiplying Eq.(A1) by the matrices $\alpha_4$ and $\Pi$, one
finds equations
\[
i\partial _t \phi(x)=\alpha_4\left[\alpha_a \partial_a+\kappa \left(\frac{1}{\xi}P_s+P_t\right)
\right]\left(\phi(x)+\chi(x)\right),~~~~~~~~~~~~~~~~~~~~~~~~~~~~~~~~~~~~~~(A4)
\]
\[
0=\left(\alpha _4^2-1 \right)\left[\alpha_a \partial_a+\kappa \left(\frac{1}{\xi}P_s+P_t\right)
\right]\left(\phi(x)+\chi(x)\right) .~~~~~~~~~~~~~~~~~~~~~~~~~~~~~~~~~~~~~~(A5)
\]
In these equations we imply that the direct sum of functions $\phi(x)$ and $\chi(x)$ is $\Psi$ ($\phi(x)+\chi(x)=\Psi$). One can verify equations
\[
\left(\frac{1}{\xi}P_s+P_t\right) \Pi =\Pi \left(\frac{1}{\xi}P_s+P_t\right)=\Pi,~~~~ \Pi\alpha_a \Pi=0,
\]
and obtain from Eq.(A5) the function $\chi(x)$:
\[
\chi(x)=-\frac{1}{\kappa}\Pi\alpha_a\partial_a\phi(x) . ~~~~~~~~~~~~~~~~~~~~~~~~~~~~~~~~~~~~~~~~~~~~~~~~~~~~~~~~~~~~~~(A6)
\]
Excluding the $\chi(x)$ from Eq.(A4) with the help of (A6) and using
the relation $\alpha_4\Pi=0$, we find the Schr\"{o}dinger equation $i\partial _t\phi (x)={\cal H}\phi (x)$ ,
with the Hamiltonian
\[
{\cal H}=\alpha_4\left[\alpha_a \partial_a+\kappa \left(\frac{1}{\xi}P_s+P_t\right) \right]
-\frac{1}{\kappa}\alpha_4\alpha_a\Pi\alpha_b\partial_a\partial_b.~~~~~~~~~~~~~~~~~~~~~~~~~~~~~~~~~~~~~~~~(A7)
\]
Although the matrices $\alpha_\mu$ are $11\times 11$-matrices, the Hamiltonian (A7) acts in $8$-dimension subspace.
One can check that Hamiltonian (A7) coincides with Hamiltonian (26). Thus, we have obtained here the Hamiltonian in terms of matrices of relativistic wave equation (7). It should be noted that Eq.(A6) is
equivalent to Eq.(21) for non-dynamical components, $\psi_{mn}$. The Schr\"{o}dinger equation with Hamiltonian (A7) does not contain non-dynamical components and can be used for solving some problems of interacting fields.

\vspace{5mm}

\textsc{\textbf{Appendix B}}

\vspace{5mm}

Now, we obtain canonical and symmetrical energy-momentum tensors and dilatation currents
starting with the second order formulation based on the Lagrangian (3).
The canonical energy-momentum tensor found from equation
\[
\Theta^c_{\mu\alpha}=\frac{\partial {\cal L}}{\partial\left(\partial_\mu A_\beta\right)}
\partial_\alpha A_\beta-\delta_{\mu\alpha}{\cal L}
\]
is given by
\[
\Theta^c_{\mu\nu}=
-F_{\mu\beta}\partial_\nu A_\beta-\xi\left(\partial_\nu A_\mu\right)\left(\partial_\alpha A_\alpha\right)
+\delta_{\mu\nu}\left[\frac{1}{4}F_{\rho\sigma}^2+\frac{\xi}{2}\left(\partial_\alpha A_\alpha\right)^2\right],~~~~(B1)
\]
and is conserved: $\partial_\mu \Theta^c_{\mu\nu}=0$.
The trace of the energy-momentum tensor is non-zero and reads
\[
\Theta^c_{\mu\mu}=\frac{1}{2}F_{\rho\sigma}^2+\xi\left(\partial_\alpha A_\alpha\right)^2.~~~~~~~~~~~~~~~~~~~~~~~~~~~~~~~~~~~~~~~~~~~~~~~~~~~~~~~~(B2)
\]
The dilatation current is given as follows \cite{Coleman}:
\[
D^c_{\mu}=x_\alpha \Theta^c_{\mu\alpha}
+\Pi_{\mu\alpha}A_\alpha,~~~~\Pi_{\mu\alpha}=
\frac{\partial {\cal L}}{\partial\left(\partial_\mu A_\alpha\right)}=-F_{\mu\alpha}-\xi\delta_{\mu\alpha} \left(\partial_\nu A_\nu\right),
\]
\vspace{-8mm}
\[
~~~~~~~~~~~~~~~~~~~~~~~~~~~~~~~~~~~~~~~~~~~~~~~~~~~~~~~~~~~~~~~~~~~~~~~~~~~~~~~~~~~~~~~~~~~~~~~~~~~~~~~~~~~~~~~~(B3)
\]
\vspace{-8mm}
\[
D^c_{\mu}=x_\alpha \Theta^c_{\mu\alpha}-F_{\mu\alpha}A_\alpha-\xi A_\mu \left(\partial_\alpha A_\alpha\right).
\]
One can check with the help of equations of motion (4) that the dilatation
current (B3) is conserved, $\partial_\mu D^c_{\mu}=0$. Thus, the
the scale invariance is valid.

The canonical energy-momentum tensor (B1) is not symmetrical.  To
obtain the symmetrical Belinfante tensor, we use the formulas \cite{Coleman}:
\[
\Theta_{\mu\alpha}^{B}=\Theta_{\mu\alpha}^{c}+ \partial_\beta X_{\beta\mu\alpha},
\]
\vspace{-8mm}
\[
~~~~~~~~~~~~~~~~~~~~~~~~~~~~~~~~~~~~~~~~~~~~~~~~~~~~~~~~~~~~~~~~~~~~~~~~~~~~~~~~~~~~~~~~~~~~~~~~~~~~~~~~~~~~~~~~(B4)
\]
\vspace{-8mm}
\[
X_{\beta\mu\nu}=\frac{1}{2}\left[\Pi_{\beta\alpha} \left(\Sigma_{\mu\nu}\right)_{\alpha\sigma}A_\sigma-
\Pi_{\mu\alpha}\left(\Sigma_{\beta\nu}\right)_{\alpha\sigma}A_\sigma-
\Pi_{\nu\alpha}\left(\Sigma_{\beta\mu}\right)_{\alpha\sigma}A_\sigma\right],
\]
where the matrix elements of the generators of the Lorentz group $\Sigma_{\mu\nu}$, in Euclidian space-time,
are given by
\[
\left(\Sigma_{\mu\nu}\right)_{\alpha\sigma}=\delta_{\mu\alpha}\delta_{\nu\sigma}
-\delta_{\mu\sigma}\delta_{\nu\alpha}.~~~~~~~~~~~~~~~~~~~~~~~~~~~~~~~~~~~~~~~~~~~~~~~~~~~~~~~~~~~~~~~~(B5)
\]
From Eq.(B4), with the help of Eq.(B5), we obtain
\[
X_{\beta\mu\nu}=F_{\mu\beta}A_\nu+\xi\left(\partial_\alpha A_\alpha\right)
\left(\delta_{\beta\nu}A_\mu-\delta_{\mu\nu}A_\beta\right),~~~~~~~~~~~~~~~~~~~~~~~~~~~~~~~~~~~~~~~~~~~~~~~~~~~~~~~(B6)
\]
so that $X_{\beta\mu\nu}$ is antisymmetric in indexes $\beta$,$\mu$, and $\partial_\beta\partial_\mu X_{\beta\mu\nu}=0$.
The symmetric and conserved Belinfante energy-momentum tensor, using Eq.(B4),(B6), becomes
\[
\Theta_{\mu\nu}^{B}=-F_{\mu\beta}F_{\nu\beta}+\xi\left(A_\nu\partial_\mu+A_\mu\partial_\nu\right)\left(\partial_\alpha A_\alpha\right)
\]
\vspace{-8mm}
\[
~~~~~~~~~~~~~~~~~~~~~~~~~~~~~~~~~~~~~~~~~~~~~~~~~~~~~~~~~~~~~~~~~~~~~~~~~~~~~~~~~~~~~~~~~~~~~~~~~~~~~~~~~~~~~~~~(B7)
\]
\vspace{-8mm}
\[
+\delta_{\mu\nu}\left[\frac{1}{4}F_{\rho\sigma}^2-\frac{\xi}{2}\left(\partial_\alpha A_\alpha\right)^2-\xi A_\beta\partial_\beta\left(\partial_\alpha A_\alpha\right)\right].
\]
At $\xi=0$ Eq.(B7) converts into energy-momentum tensor of electrodynamics. Energy-momentum tensor similar to (B7) was found in \cite{Jose2} by varying action on the metric tensor. A modified dilatation current is given by \cite{Coleman}
 \[
D_\mu^B=x_\alpha \Theta_{\mu\alpha}^{B}+V_\mu,~~~~~~~~~~~~~~~~~~~~~~~~~~~~~~~~~~~~~~~~~~~~~~~~~~~~~~~~~~~~~~~~~~~~(B8)
\]
where the field-virial $V_\mu $, in our case, becomes
\[
V_\mu=\Pi_{\mu\alpha} A_\alpha-\Pi_{\alpha\beta} \left(\Sigma_{\alpha\mu}\right)_{\beta\sigma}A_\sigma=
2\xi A_\mu\left(\partial_\alpha A_\alpha\right).~~~~~~~~~~~~~~~~~~~~~~~~~~~~~~~~~~~~(B9)
\]
One can obtain the trace of the Belinfante tensor (B7):
\[
\Theta_{\mu\mu}^{B}=-2\xi \partial_\mu \left[A_\mu\left(\partial_\alpha A_\alpha\right)\right].~~~~~~~~~~~~~~~~~~~~~~~~~~~~~~~~~~~~~~~~~~~~~~~~~~~~~~~~~~~~(B10)
\]
From Eq.(B8),(B10), we obtain
\[
\partial_\mu D_\mu^B=\Theta_{\mu\mu}^{B}+\partial_\mu V_\mu=0,
\]
and the modified dilatation current is conserved. Thus, the scale invariance takes place with the conserved currents (B3) and (B8).

\vspace{5mm}

\textsc{\textbf{Appendix C}}

\vspace{5mm}

Let us consider the quantization of fields for given Lagrangian (3) (see also \cite{Itzykson}).
Conjugated momenta for generalized ``coordinates" $A_\mu (x)$ are given by
\[
\pi_0(x)=\frac{\partial {\cal L}}{\partial \dot{A}_0}=-\xi \partial_\mu A_\mu,~~~~~~
\pi_m(x)=\frac{\partial {\cal L}}{\partial \dot{A}_m}=\dot{A}_m+\partial_m A_0.~~~~~~~~~~~~~~~~(C1)
\]
Then the density of the Hamiltonian becomes
\[
 H=\pi_m \dot{A}_m-\pi_0 \dot{A}_0-{\cal L}=\dot{A}_m\left(\dot{A}_m+\partial_m A_0\right)
\]
\vspace{-8mm}
\[
~~~~~~~~~~~~~~~~~~~~~~~~~~~~~~~~~~~~~~~~~~~~~~~~~~~~~~~~~~~~~~~~~~~~~~~~~~~~~~~~~~~~~~~~~~~~~~~~~~~~~~~~~~~~~~~~(C2)
\]
\vspace{-8mm}
\[
+\frac{1}{4}F_{\mu\nu}^2-\xi \dot{A}_0\partial_\mu A_\mu+\frac{1}{2}\xi\left( \partial_\nu A_\nu\right)^2.
\]
One can verify that the equality $ H=\Theta^c_{44}$ holds where the canonical energy-momentum tensor
$\Theta^c_{\mu\nu}$ is given by (B1). It should be noted that the classical energy ${\cal E}=\int Hd^3x$ is not bounded from below and the system is unstable. Therefore, we need to introduce indefinite metrics for quantization. With the help of standard commutation relations for canonical variables
$\left[A_\mu (\textbf{x},t),\pi_\nu (\textbf{y},t) \right]=i\delta_{\mu\nu}\delta\left(\textbf{x}-\textbf{y}\right)$,
one obtains
\[
\left[A_n (\textbf{x},t), \dot{A}_m (\textbf{y},t)+ \partial_m A_0 (\textbf{y},t)\right]=i\delta_{mn}\delta\left(\textbf{x}-\textbf{y}\right),
\]
\vspace{-8mm}
\[
~~~~~~~~~~~~~~~~~~~~~~~~~~~~~~~~~~~~~~~~~~~~~~~~~~~~~~~~~~~~~~~~~~~~~~~~~~~~~~~~~~~~~~~~~~~~~~~~~~~~~~~~~~~~~~~~(C3)
\]
\vspace{-8mm}
\[
\left[A_0 (\textbf{x},t),\partial_\mu A_\mu (\textbf{y},t) \right]=-\frac{i}{\xi}\delta\left(\textbf{x}-\textbf{y}\right).
\]
In the momentum space the real fields $A_\mu$ read
\[
A_\mu (x)=\sum_\textbf{k}\frac{1}{\sqrt{2Vk_0}}\left[a_\mu (\textbf{k})e^{ik_\mu x_\mu} +a^+_\mu (\textbf{k})e^{-ik_\mu x_\mu}\right],~~~~~~~~~~~~~~~~~~~~~~~~~~~~~~~~~~~~~~~~(C4)
\]
where $k_\mu^2= \textbf{k}^2-k_0^2=0$, $V$ is the normalization volume. It should be noted that the field $A_\mu$ possesses four independent components: two components are transverse, one component is longitudinal, and one component corresponds to the scalar polarization. In \cite{Kruglov}, \cite{Kruglov1}, we found four independent solutions to equations of motion in the form of matrix-dyads for the gauge $\xi=1$. The fields (C4) satisfy commutators (C3) if creation and annihilation operators
obey the commutation relations as follows:
\[
\left[a_m(\textbf{k}),a^+_n(\bar{\textbf{k}})\right]=\delta_{mn}\delta\left(\textbf{k}-\bar{\textbf{k}}\right),~~~~
\left[a_0(\textbf{k}),a^+_0(\bar{\textbf{k}})\right]=-\frac{1}{\xi}\delta\left(\textbf{k}-\bar{\textbf{k}}\right).~~~~~~(C5)
\]
 For the Feynman gauge $\xi =1$ (which was used in \cite{Kruglov1}) the RWE (6), (10) (and (A7)) are simplified. In this case the ``wrong" sign ($-$) in the commutator for $a_0(\textbf{k})$, $a^+_0(\bar{\textbf{k}})$ in (C5) indicates on the necessity of introducing the indefinite metrics. For any gauge $\xi$ there are difficulties with the presence of the ghosts if one considers four polarizations of the field $A_\mu$ to be physical. In QED the photon fields possess only two polarizations and physical values do not depend on the gauge $\xi$ due to the restriction on the physical Hilbert space \cite{Itzykson}.

\end{document}